\documentstyle[editedvolume,namedreferences,epsf]{crckapb} 

\def\etal   {~{\it et~al.}}

\def\kms    {\ km\,s$^{-1}$}
\def\pix    {\ pix$^{-1}$}
\def\cmsq   {\ cm$^{-2}$}

\def\HI     {$\lambda$21cm}
\def\Ha     {\hbox{H$\alpha$}} 
\def\hi     {\hbox{H\,{\sc i}}} 
\def\hii    {\hbox{H\,{\sc ii}}} 
\def\nii    {\hbox{[N\,{\sc ii}]}} 
\def\deg    {\hbox{${}^\circ$}}
\def\arcsec {\hbox{${}^{\prime\prime}$}}

\def\farcs  {\hbox{$.\!{}^{\prime\prime}$}}

\begin{opening}

\title{KINEMATICS AND MASS MODELLING OF NGC 1068}

\author{Walter Dehnen}           \institute{Theoretical Physics, Oxford}
\author{Jonathan Bland-Hawthorn} \institute{Anglo-Australian Observatory}
\author{Andreas Quirrenbach}     \institute{Max-Planck-Institut f\"ur
                                            Extraterrestrische Physik} 
\author{Gerald N.~Cecil}         \institute{University of North Carolina}

\date{December 1996}

\end{opening}

\runningauthor  {W.~Dehnen\etal}

\runningtitle   {Kinematics and Mass Modelling of NGC 1068}

\begin{document}

\begin{abstract}
  We present the kinematics of the ionized gas over the inner 140\arcsec\ (10 
  kpc) from observations with the HIFI Fabry-Perot interferometer. There is
  clear evidence for density wave streaming and bar-driven streaming across 
  the field, with bi-symmetric arms that penetrate to within 200 pc of the
  nucleus. CO maps show linear structures along (although slightly offset from)
  the bar consistent with a strong shock. Along the spiral arms which encircle 
  the bar, the \hii\ regions lie downstream of the CO gas in the rest frame of
  the bar, as do the dust lanes, only if the gas outruns the stellar bar.
  As a first step towards understanding the details of the gas kinematics, and
  attempting to determine the mass inflow rate towards the nucleus, we build a
  mass model for the central disk constrained by near-infrared images. We plan
  to use this model as gravitational background potential for hydro-dynamical
  simulations of the gas response to the bar. Comparing these with the data 
  presented should enable us to constrain various quantities such as pattern 
  speed, stellar mass-to-light ratio, central mass concentration, and gas 
  fueling rate.
\end{abstract}
\newpage


\begin{figure}
 \centerline{\epsfxsize=6.2truecm \epsfbox{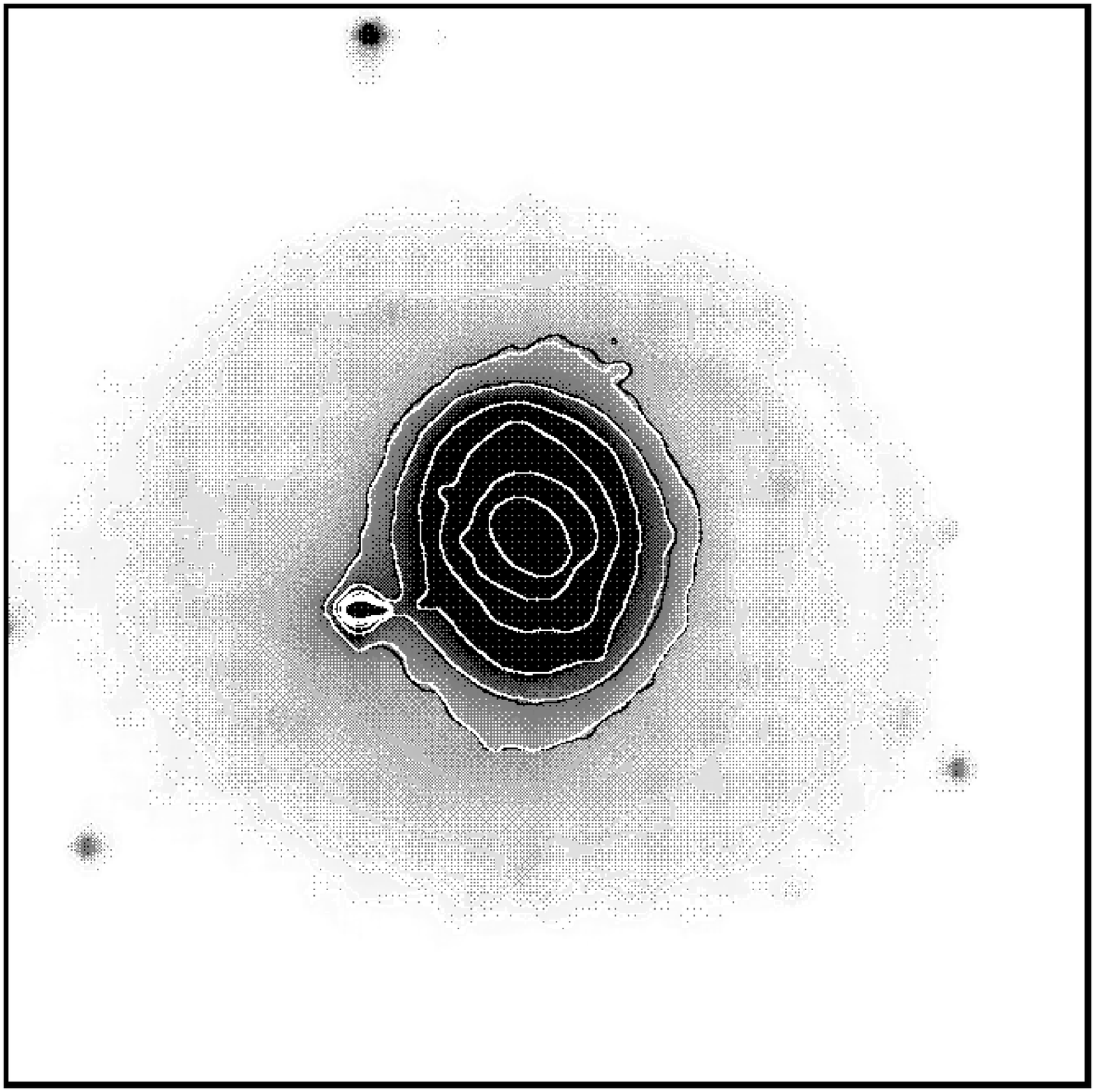}
             \epsfxsize=6.2truecm \epsfbox{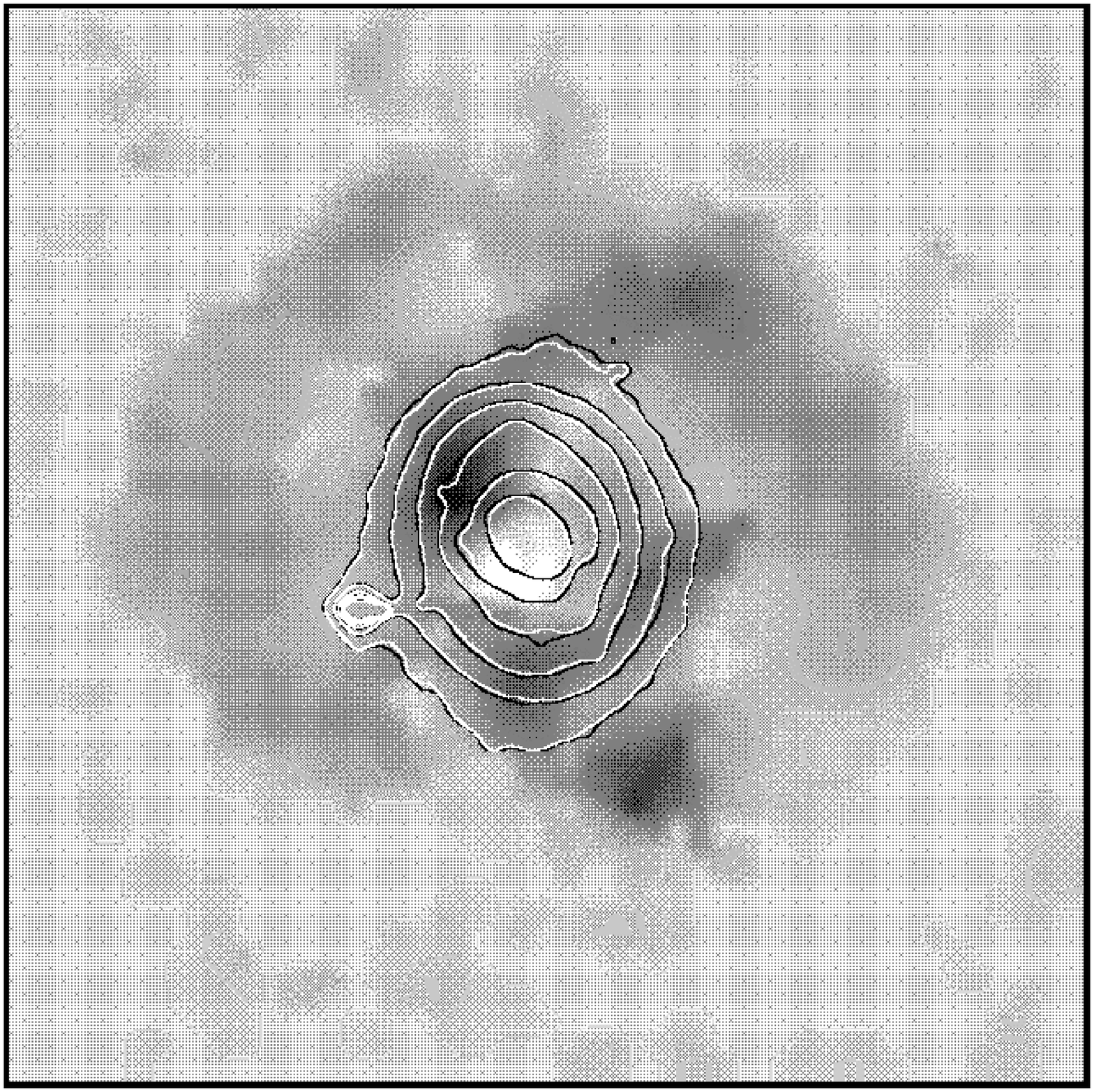}}
 \caption[]{
  {\it Left:} (a) $B$ band image of the outermost disk superimposed on $I$ band
  contours. The field of view is 500\arcsec\ (37.5 kpc) sampled at 10\arcsec
  \pix. {\it Right:} (b) VLA \HI\ observations (beam size $\approx$ 50\arcsec\
  FWHM) of the distribution of atomic hydrogen throughout NGC 1068. The peak
  column density is 1.2$\times$10$^{21}$\cmsq; the column density falls below
  10$^{20}$\cmsq\ beyond a radius of $\approx$18 kpc. Note the prevalent
  absorption in the nuclear regions, the dominant northern arm and the outer
  ring of hydrogen. $I$ band isophotes have been superimposed (contoured) at
  equal log intervals with a resolution of $8\arcsec$ FWHM. A clear oval
  distortion is seen with PA $\approx0\deg$. The SE source is a saturated star.}
 \label{oval}
\end{figure}

\section{Introduction}
The luminous Seyfert 2 galaxy NGC 1068\footnote{
        Unless otherwise stated, global quantities for NGC 1068 are taken from
        the Ringberg Standard \cite{BH97}.}
lies at the forefront of attempts to unify the broader class of ``active 
galaxies'' within a single physical framework \cite{AM85}. Thus, it is 
surprising that so little is known of the large-scale gas motions in this
nearby galaxy and, in particular, their connection with the nuclear activity.

\section{Gas Kinematics}
\subsection{\Ha\ Observations}
NGC 1068 was observed on the night of Dec.~8, 1985 using the Hawaii Imaging 
Fabry-Perot Interferometer (HIFI) at the CFH 3.6m telescope.  The observations
were made using an etalon with 85\AA\ free spectral range and a reflective 
finesse of 60. NGC 1068 was observed through two 50\AA\ FWHM blocking filters 
with central wavelengths $\lambda$6615 and $\lambda$6585 to cover \Ha\ and the
neighbouring lines of \nii$\lambda\lambda$6548,6583.  
Observations were also taken for the \nii\ lines over the period Dec.~5-7,
1985 but cirrus and variable weather conditions have compromised these data. 
The Fabry-Perot system is placed at the f/8 Cassegrain focus and has an output
beam of f/2 after focal reduction. At the time of the observations, the best 
available imaging detector was the TI $512\times512$ 3-phase CCD whose 15$\mu$
pixels subtend 0.43\arcsec\ on the sky giving a $220\arcsec\times220\arcsec$
field-of-view. The data have a velocity resolution of 65\kms\ FWHM.
The \Ha\ and blue \nii\ lines were sampled at 40\kms, while the red \nii\ line
was sampled more coarsely at 78\kms. The etalon produces a total velocity 
coverage of $\sim$4000\kms\ free of order confusion although only half this 
range was observed. The spatial resolution of the data is 1\arcsec\ in regions
of high signal-to-noise ratio (e.g. the starburst ring), and roughly 2\arcsec\
in the diffuse gas.

\begin{figure}
 \centerline{\epsfxsize=6.2truecm \epsfbox[62 120 558 629]{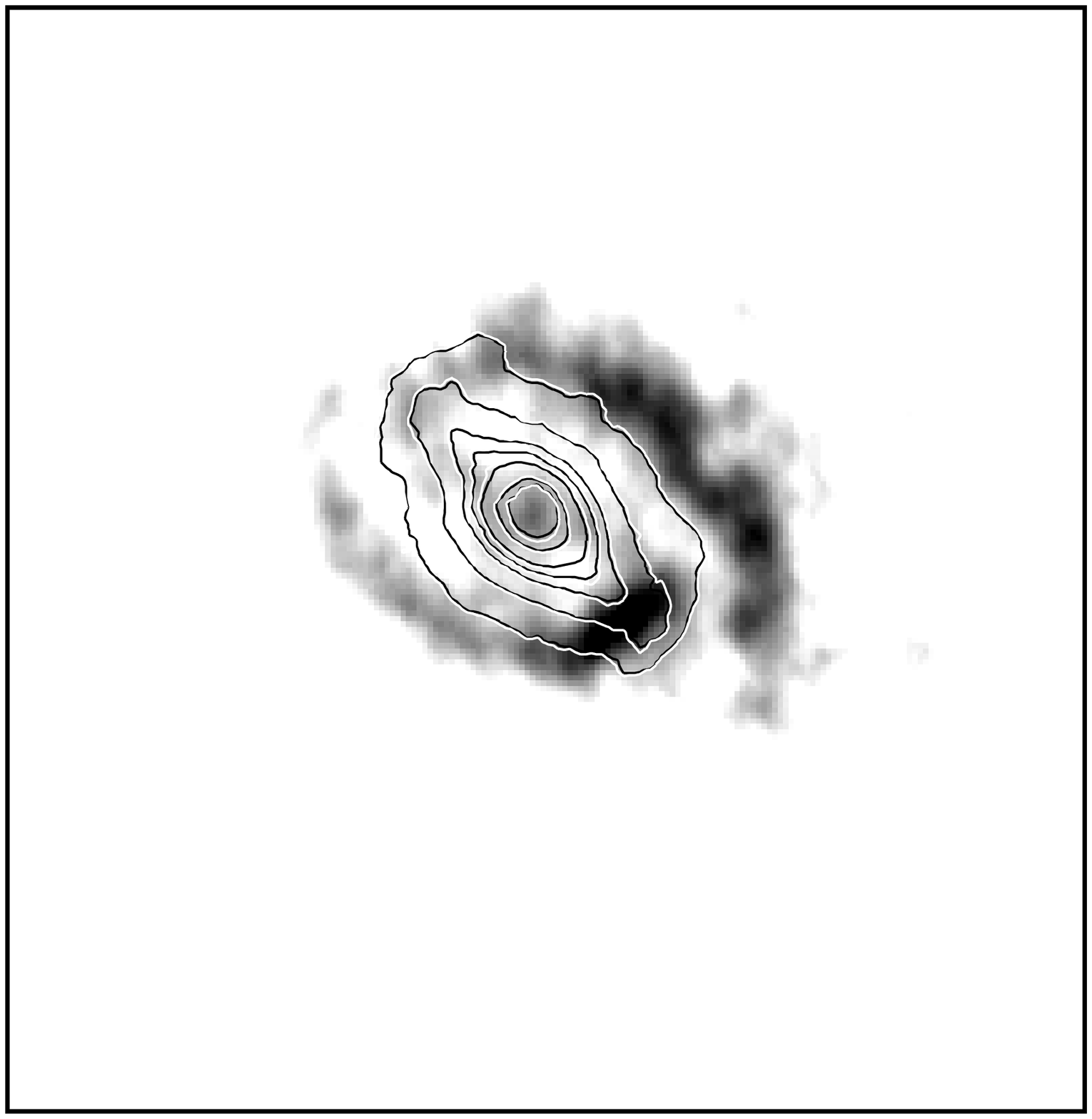}
             \epsfxsize=6.2truecm \epsfbox[60 120 556 629]{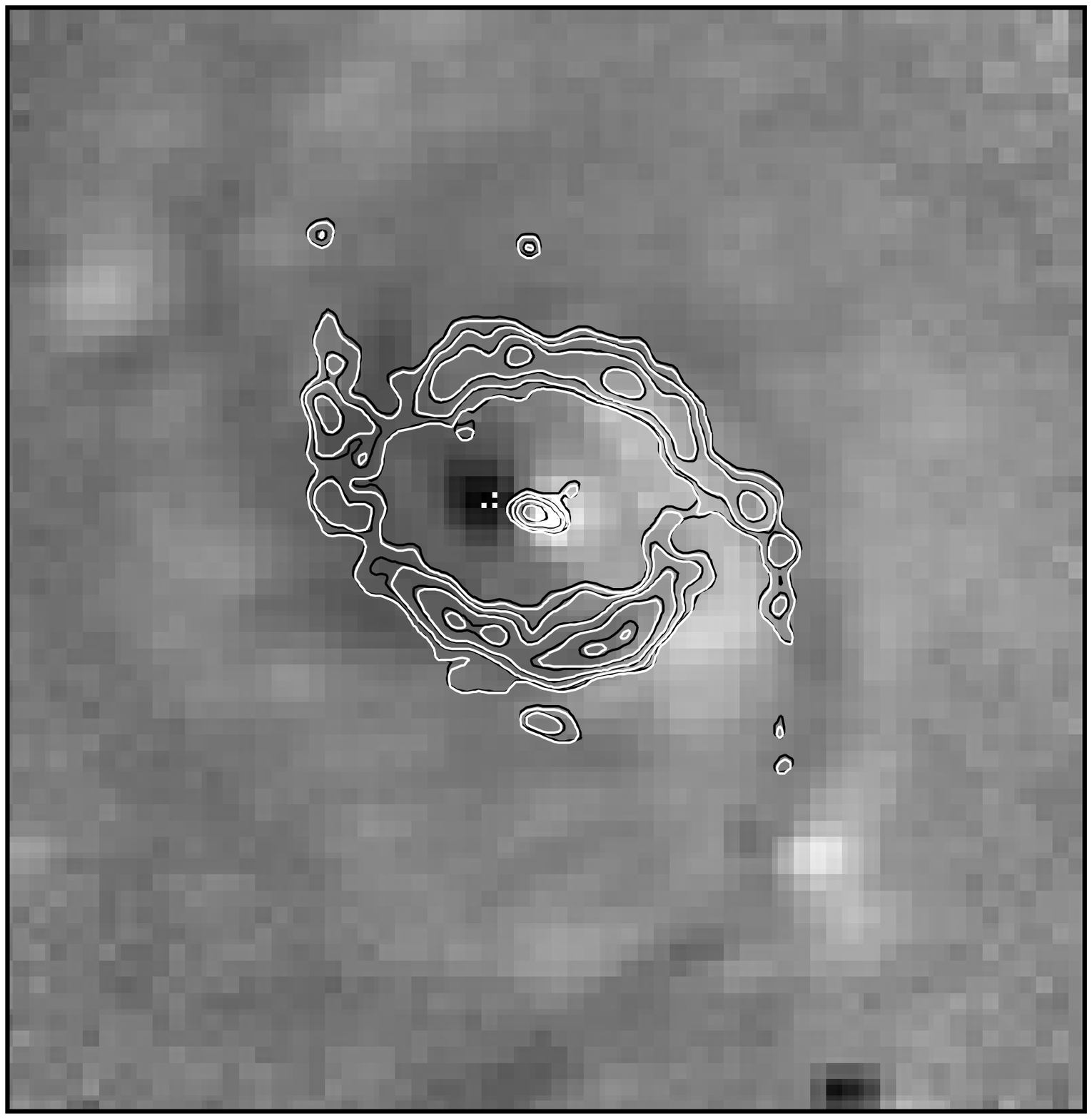}}
 \caption[]{
  {\it Left:} (a) CO emission \cite{HB95} superimposed onto the $K$ band image
  of NGC 1068 \cite{THR89} over a field of 100\arcsec. The CO gas distribution
  is bi-symmetric about the nucleus, particularly along the bar, where it sits 
  east of the bar axis to the northeast, and west of the bar axis to the
  southwest. 
  {\it Right:} (b) CO emission \cite{PLA91} superimposed onto a B$-$I color
  image to emphasize that the cold gas sits {\it along the inside} of the
  dominant spiral pattern delineated by the optical arms \cite{SAN61}. This
  is particularly clear to the SW. The dark dust lane and optical arm correspond
  to B$-$V$\approx$1.0 and B$-$V$\approx$0.5 respectively \cite{ICH87}.}
\label{bar}
\end{figure}

\begin{figure}
 \centerline{\epsfxsize=6.2truecm \epsfbox{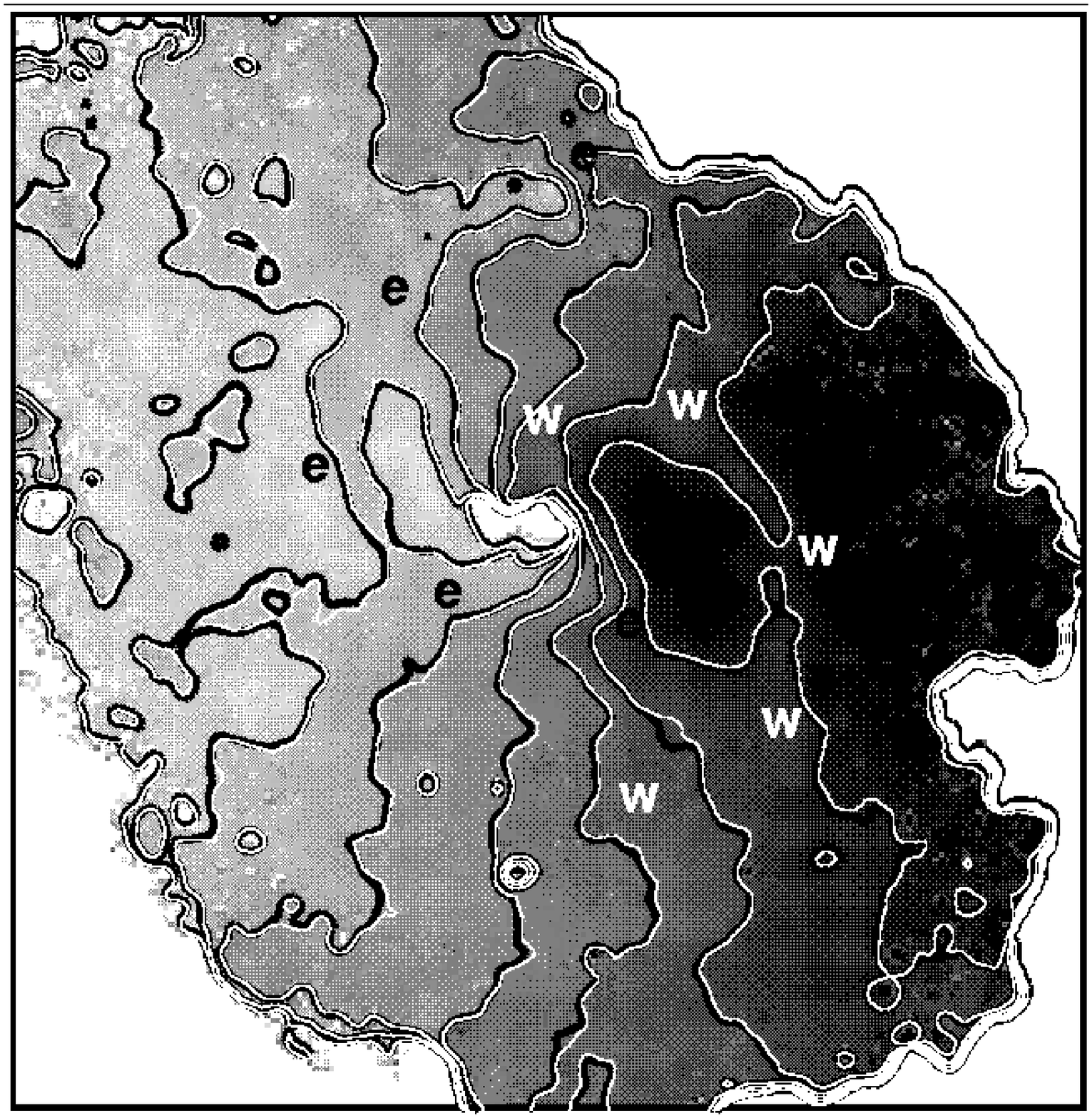}
             \epsfxsize=6.2truecm \epsfbox[62 120 558 629]{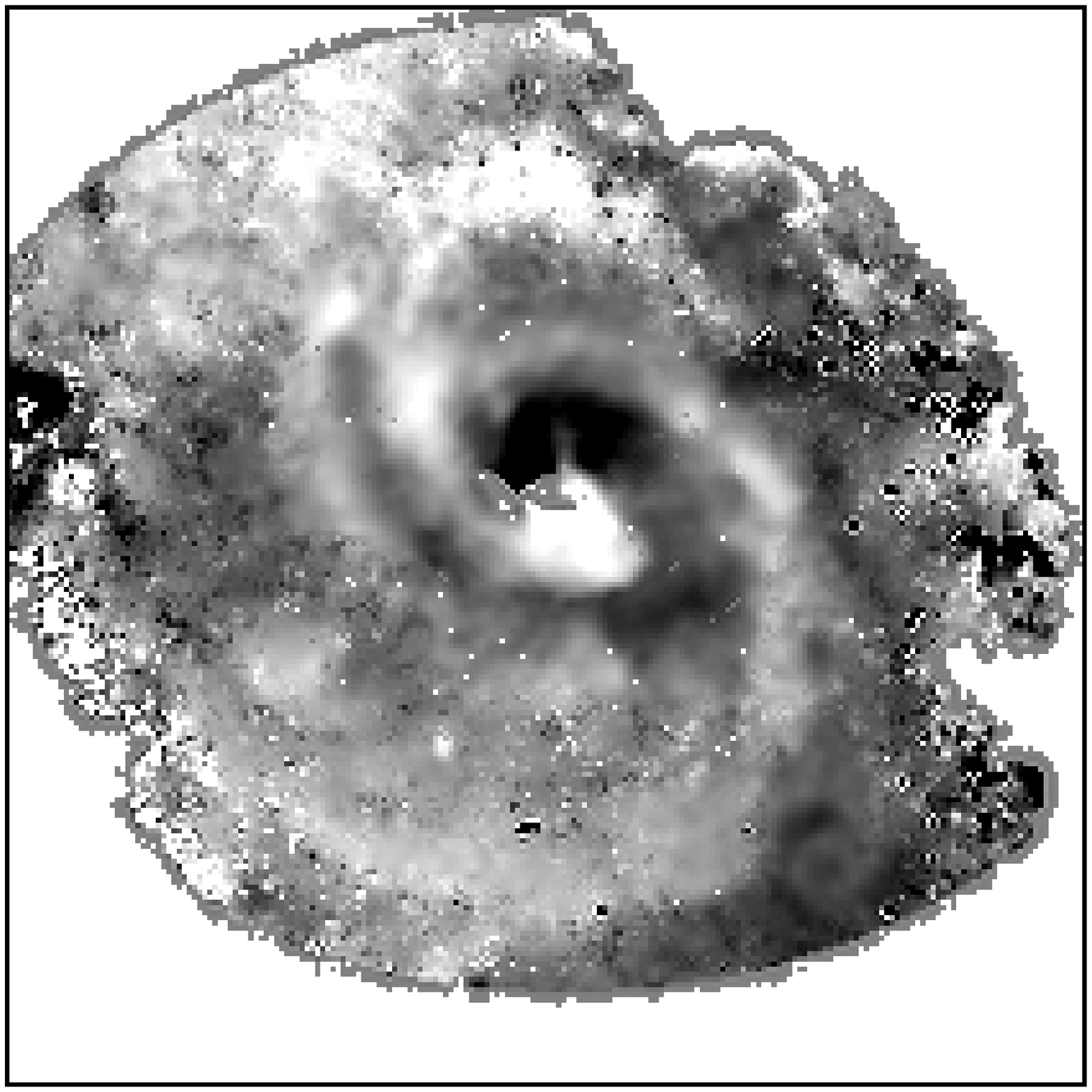}}
\caption[]{
  {\it Left:} (a) \Ha\ velocity field over the same field as Fig.~\ref{bar}. 
  The contours from left to right correspond to --125, --75, --25, 25, 75, and
  125\kms\ with respect to systemic velocity. Within the inner 40\arcsec, the
  Z-shaped contours, characteristic of streaming along the bar, are well 
  pronounced. This is particularly clear to the SW where the extended 
  narrow-line region is much fainter and therefore does not pollute the bar
  streaming. The ``w'' and ``e'' labels show how the western and eastern 
  kinematic arms spiral out from the nucleus into the large-scale disk. The
  central hole arises from problems in line fitting within the extended narrow 
  line region; the AGN falls at the western edge of this hole.
  {\it Right:} (b) The residual velocities after subtracting a rotating disk
  model produced
  from the \Ha\ rotation curve in Fig.~\ref{gasrot}. Streaming motions are seen
  over the entire field, being particularly strong along the bar and in the two 
  kinematic ``e'' and ``w'' arms. The greyscale range is the same as in (a). }
\label{kinematics}
\end{figure}

\begin{figure}
 \centerline{ \epsfxsize=6.2truecm \epsfbox[60 120 570 629]{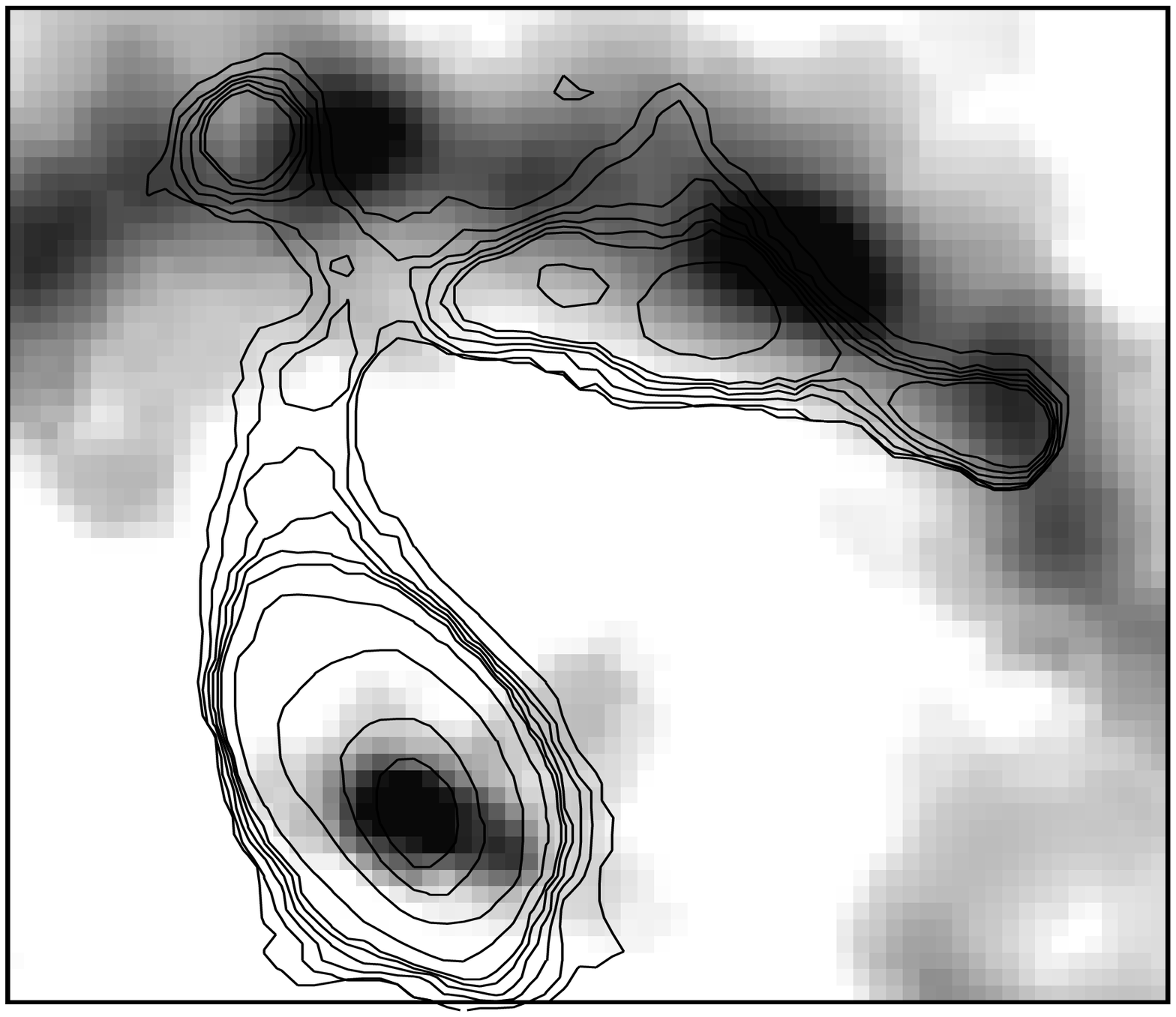}}
  \caption{ The \Ha\ flux map contoured over on the CO emission for a 
  25\arcsec\ field north of the nucleus. The ``w'' feature in (a) shows 
  continuity in velocity along this feature, even across the CO spiral arms.
  Density wave action can be traced all the way into the narrow line region NE
  of the nucleus.}
\label{streamer}
\end{figure}

\subsection{Large-Scale Structure}
 Deep optical and \hi\ images of NGC 1068 (Fig.~\ref{oval}) show an outer
$\theta-$shaped ring and a bright central disk with diameter $\sim$20 kpc
oriented perpendicular to the outer ring. The innermost $I$ band contour has
the same dimension as the stellar bar seen at $K$ (Fig.~\ref{bar}). In 
Fig.~\ref{kinematics}, we present the velocity field for the ionized
gas over the inner disk.  The velocity field is shown for a field of view of
100\arcsec\ and reveals a series of concentric ``kinematic spiral arms'' 
which appear superimposed on ordered, large-scale rotation. As shown in
Fig.~\ref{gasrot}, the large-scale disk appears to undergo flat rotation 
beyond a radius of 1.5 kpc with a kinematic major axis that is consistent with
the photometric position angle of the outer ring. Across the inner 40\arcsec,
the Z-symmetric velocity contours are the hallmark of highly radial orbits
along a bar (e.g.\ Huntley 1978). Thus, the observed elliptic streaming is 
consistent with the presence of the stellar bar detected at near-infrared
wavelengths \cite{SCO88,THR89}.

\subsection{Observed Kinematics}
In Fig.~\ref{kinematics}$a$, the velocity field of the ionized gas is derived 
from the weighted sum of the \Ha\ and \nii\ kinematics. NGC 1068 has probably 
the most completely observed velocity field of any barred spiral galaxy due to 
a `diffuse ionized medium' which pervades the inner disk \cite{BSC91}. The 
large-scale disk clearly undergoes ordered rotation. However, notice the 
bisymmetric streaming features $25^{\prime\prime}$ to the east and west of the
nucleus, which are particularly apparent in the residual velocity field
(Fig.~\ref{kinematics}$b$) after subtracting a rotating disk model. These
are retarded with respect to the circular motion by 10\kms\ on average
reaching 30\kms\ in places, which probably corresponds to 30\kms\ up to 50\kms\
in the plane of the galaxy. Additional kinematic spiral anomalies at larger 
radius are more diffuse. The characteristic Z-symmetry of an oval distortion 
(cf.\ Fig.~14 of Roberts\etal, 1979) can be traced all the way into the nucleus
from the southwest, where the southeastern spiral feature is less noticeable. 
To the north, the spiral feature exhibits continuity in velocity as it crosses 
the elliptic ring. The western kinematic feature has a counterpart in the \Ha\
flux map (Fig.~\ref{streamer}).

\begin{figure}
 \epsfxsize=\textwidth \epsfbox{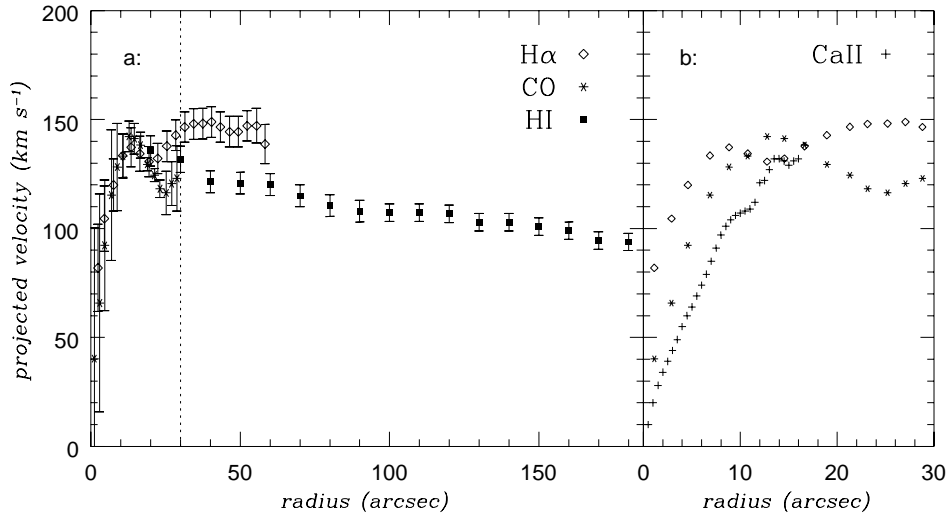}
\caption[]{(a) Gas rotation curve observed in \Ha, CO and \hi. (b) The points 
inside the dashed line in (a) have been blown up to show how \Ha\ and CO 
compare with the Ca~II stellar velocities from Garc\'{\i}a-Lorenzo\etal\ (1997).
The absorption line measurements have typical uncertainties of $\pm$20\kms.}
\label{gasrot}
\end{figure}

The \Ha, \hi\ (21~cm), and $^{12}$CO ($J{=}1\rightarrow 0$) rotation curves are
presented in Fig.~\ref{gasrot}. There is good agreement between the CO 
\cite{HB95} and the \Ha\ kinematics although rather less agreement with the \hi\
data (VLA observations supplied by H.~Liszt). However, the \hi\ data use a 
15\arcsec\ beam and the data have yet to be corrected for beam smearing and 
absorption in the centre. The velocity field shows extensive evidence for strong
density wave streaming on all scales. In the \Ha\ data, there is a sharp 
transition region between 0\arcsec\ and 10\arcsec\ where the PA changes smoothly
between 225$\deg$ and 80$\deg$ \cite{ARR96}.  A similar effect is seen in the 
\hi\ data between 70\arcsec\ and 100\arcsec\ where the PA changes smoothly from
91$\deg$ to 106\deg. For the \Ha\ and CO ring fits, the adopted inclination was
40$\deg$ although the orbit PA was a free parameter.

By subtracting the simple model of elliptic streaming overlayed on flat 
rotation, we have discovered bisymmetric spiral streamers in both the ionized
line flux and kinematics that penetrate from $\sim$1 kpc to within 100 pc of
the AGN.  We find evidence for this effect in both the CO line flux and
kinematics, albeit at lower spatial resolution \cite{KAN89}. By implication, it
is likely that the observed non-circular motions are confined to the galactic
plane and arise from bar-driven density waves. In collaboration with Harvey 
Liszt (National Radio Astronomical Observatory), we find tentative evidence 
that, while the $\lambda$21cm and CO kinematics are largely consistent, the
\Ha\ kinematics show substantial deviations from the cold phase at many 
positions within the inner disk away from the spiral streamers.

\section{Mass Model}
\subsection{Why we need a Mass Model}
Several studies of the gaseous response to rotating bar potentials can be found
in the literature (e.g., Huntley\etal, 1978; Athanassoula, 1992). One of the 
main results is the occurrence of shocks at characteristic positions. One of 
these is an often tightly wound, trailing, and bisymmetric spiral starting at
both ends of the bar, exactly as observed for NGC 1068 (Fig.~\ref{bar}$b$).
Associated with the shock is dust and cool gas visible in CO emission 
as spiral arms in Fig.~\ref{bar}$a$. Another place where shocks typically
appear is along the bar usually somewhat offset from the major axis in the
leading direction. For NCG 1068, Helfer \& Blitz (1995) detected CO emission 
along the bar which indeed appears offset from the stellar bar's major axis 
towards the leading edge (Fig.~\ref{bar}$a$). Moreover, the models often
have a central concentration of gas orbiting on near-circular, so-called
x2-orbits, also evident in CO observations of NGC 1068. Thus, {\em
qualitatively}, this fits nicely into the overall picture of the bar rotating
prograde with respect to the disk. 

However, theoretical studies also revealed a sensitive dependence of position
and strength of the shocks to overall quantities such as pattern speed, bar
strength, central mass concentration, and sound speed of the gas \cite{ATH92,%
ENG97}. For a {\em quantitative\/} understanding of the gaseous morphology and
kinematics in NGC 1068, a more detailed study is therefore highly desirable 
incorporating all information that is available on those quantities. In 
particular, the NIR light ought to be a good tracer of the underlying (stellar)
mass distribution.

\subsection{Creating a Mass Model}
Using NIR light as mass tracer to construct the gravitational potential for 
barred galaxies is by now a common technique (e.g., Quillen\etal, 1994). A 
little complication, which arises in the case of NGC 1068, is the presence of a
central point source due to dust heated by the central AGN and emitting in the
NIR. Here, we used three $K$-band images, (a) a very high-resolution (0\farcs2
FWHM) image of the centre, (b) a high-resolution image (0\farcs4 FWHM) of the
stellar bar, and (c) a low resolution image (2\arcsec\ FWHM) of the whole disk. 
The latter of these was supplied by H.~Thronson \cite{THR89}, while images a
and b where obtained with SHARP at the NTT (3.5m) using short exposures (0.1
and 0.5 seconds, respectively) in conjunction with a shift-and-add technique
\cite{QUI97}.

A scaled PSF (an image of a star obtained under the same conditions) accounting
for the point source was subtracted from image a, of which subsequently the
azimuthally averaged luminosity profile was evaluated. To correct image b for
the central point source, it was first split into non-axisymmetric and
axisymmetric (azimuthally averaged) parts; second, inside 2\farcs8 the latter
was replaced by the profile obtained from image a; and finally, these were added
back to the non-axisymmetric part. This procedure gave a high-resolution map of
the stellar NIR emission inside 30\arcsec. Finally, the map was extended with 
image c (though with lower resolution) to $160\arcsec\times160\arcsec$
corresponding to $11.5{\rm kpc}\times11.5{\rm kpc}$.

\begin{figure}
 \centerline{ \epsfxsize=6.2truecm \epsfbox{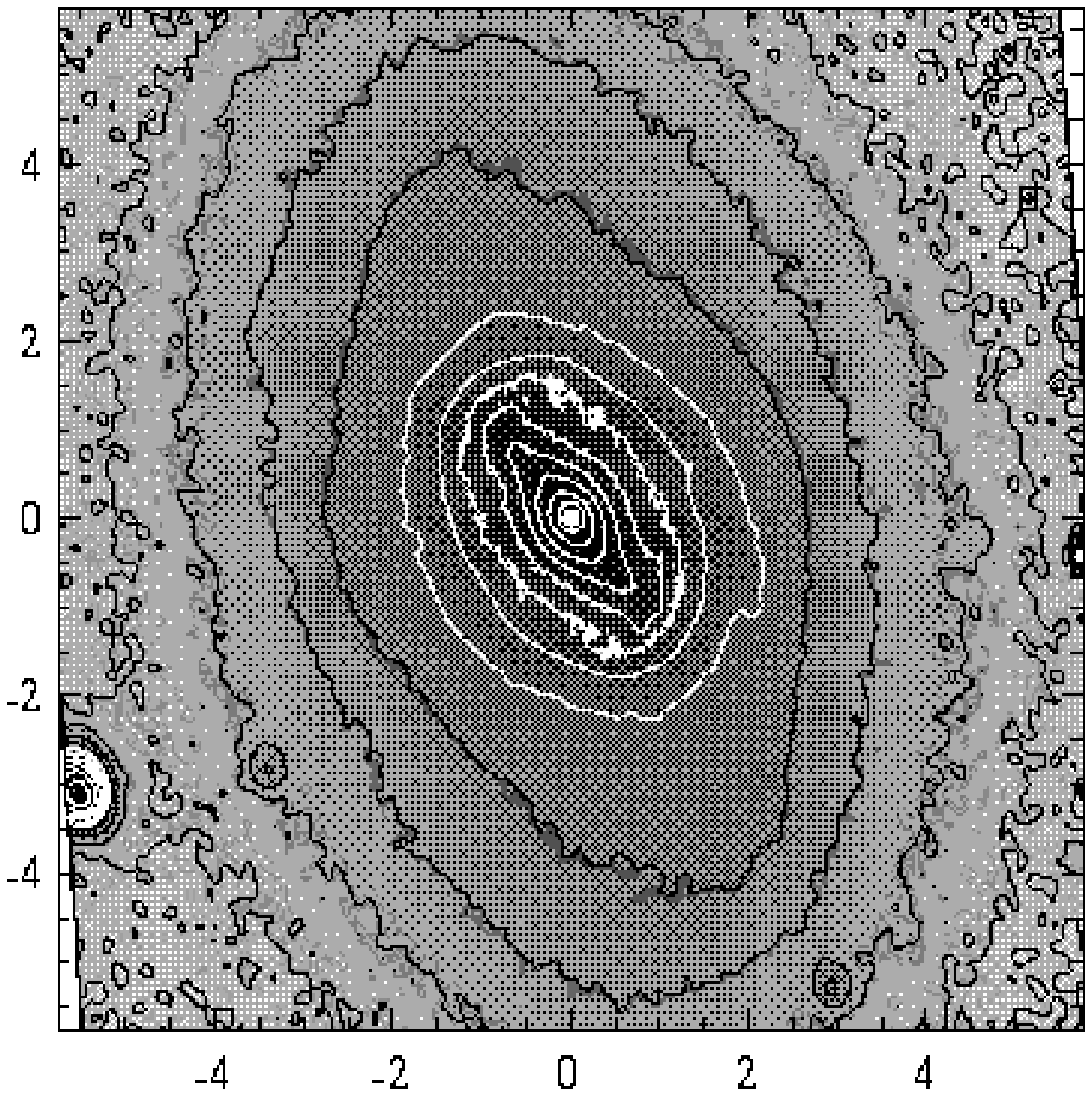}
              \epsfxsize=6.2truecm \epsfbox[70 196 570 550]{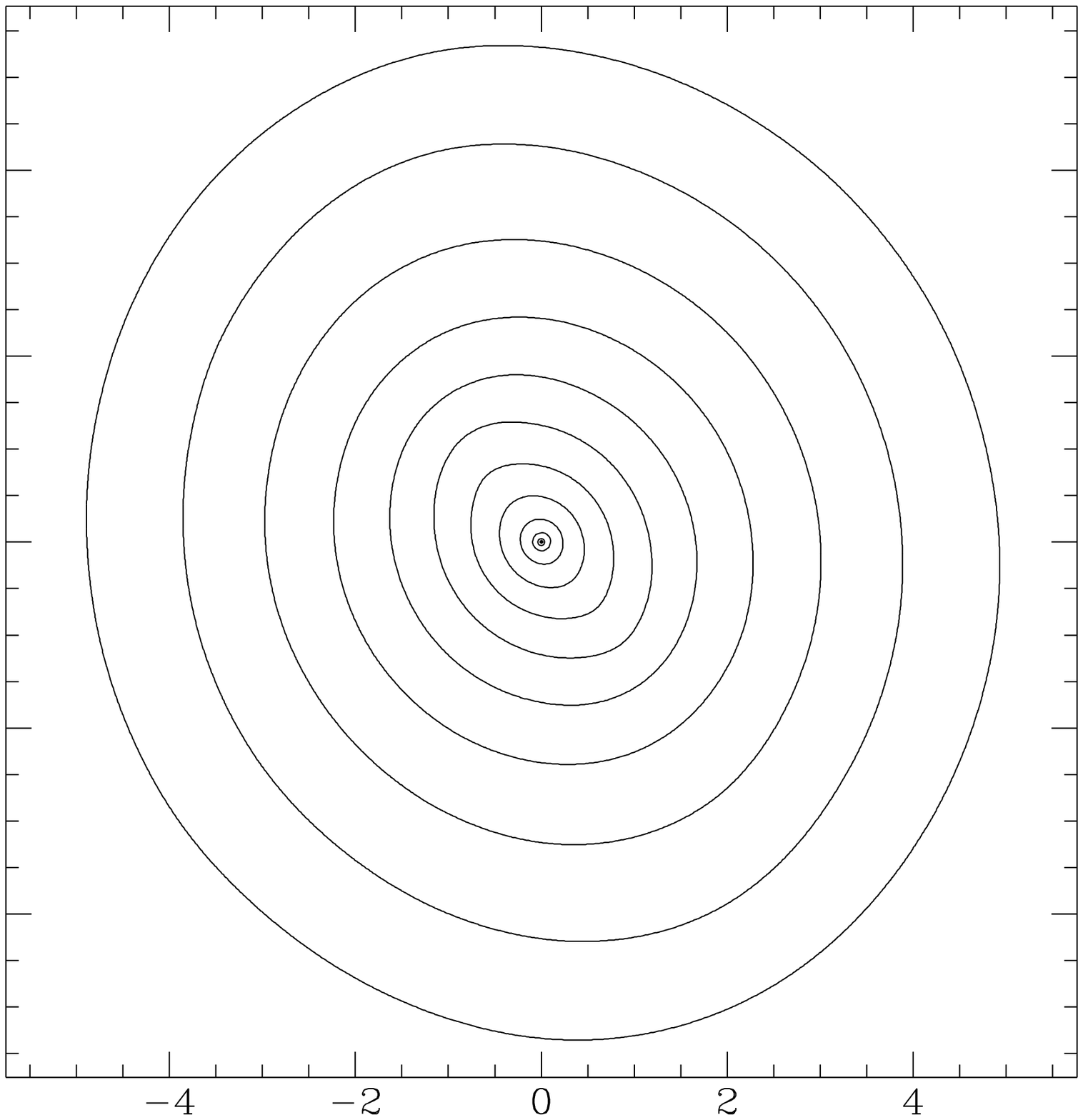}}
 \caption[]{{\it Left:} (a) Density of our model (disk \& bulge) in the plane
  of the galaxy ($z=0$). Contours are spaced by 0.25 dex. Labels are in kpc.
  The bright star in the south-east does not enter the evaluation of the 
  potential. Note that the angle between the major axes of inner bar and outer
  oval is smaller than it appears in projection (Fig.\ \ref{oval}). 
  {\it Right:} (b) Isocontours of the gravitational potential created by the
  density of disk and bulge, a black hole has not been added. Contours
  are spaced by 0.1 dex. The potential is both smoother and rounder than the
  mass distribution.}
 \label{mass-model}
\end{figure}

From this map a mass model was derived as follows:
(i)   a spherical model for the central (apparently round) bulge was subtracted;
(ii)  the remaining disk was de-inclined assuming PA=106\deg\ and $i$=40\deg\
      \cite{BH97};
(iii) the resulting surface density was vertically thickened by the profile
      $(2h)^{-1}\exp(-|z|/h)$ with $h$=200 pc; and
(iv)  the spherical model for the central bulge was added back.
Similarly, dark mass components such as a central black hole\footnote{
	Presumably, in studies of the gas kinematics, a central point mass
	can hardly be distinguished from a very dense star cluster. However,
	the latter possibility might be ruled out by physical arguments
	(e.g.\ lifetime of such a cluster). Additional constraints on the
	central potential are given by maser kinematics \cite{GRE96}, though
	their interpretation is not straightforward.}
and a halo may be added. The scale height of 200 pc is in agreement with a 
ratio of ${\sim}0.1$ between scale height and length typical for disk galaxies
in the infrared \cite{WAI89,BAR92}.

Clearly, many parameters entering the above procedure are a priori not very 
well determined, e.g., the scale height and black hole mass. In particular,
the dominance of the hot dust in the centre causes considerable uncertainty
in the central stellar light distribution. Here we have used a $\gamma$-model
\cite{DEH93} for the bulge component, the volume density of which, 
\begin{equation}
	\rho_\gamma(r) = {{\rm M}(3-\gamma) \over 4\pi}
			 {r_0 \over r^\gamma (r+r_0)^{4-\gamma}},
\end{equation}
is proportional to $r^{-\gamma}$ for $r$ much smaller than the scale radius
$r_0$ and falls off as $r^{-4}$ at large radii.
Our parameters, $\gamma=1.5$ and $r_0=0\farcs7$, are not well constrained
by the $K$ band photometry. For the future, we plan to improve on this point by
taking into account also the $J$ and $H$ band which are less sensitive to 
emission from the hot dust in the centre. These data, in combination with the
$K$ image, may also be used to correct for dust absorption along the bar, which
can be a problem particularly near the shocks.

The volume density in the plane of the galaxy resulting from this procdure
is shown in Fig.~\ref{mass-model}$a$. As can be seen from the spacing of the
contours, the scale length of the central bar is 
significantly smaller than that of the outer oval. 

For gas kinematics and hydro-dynamical simulations, only the potential (and
forces) in the plane $z\,{=}\,0$ are of interest. For the disk this has been
evaluated from the expansion of the density in azimuthal harmonics up to order
$m=16$. The gravitational potential of the bulge model is given by
\begin{equation}
  \Phi_\gamma(r) = - {G{\rm M}\over r_0} \times \left\{
  \begin{array}{ll}
	\displaystyle {1\over2-\gamma} \left[1 - 
		\left({r\over r+r_0}\right)^{2-\gamma}\right] \qquad
	& \gamma\neq2 \\ & \\
	\displaystyle \ln{r\over r+r_0} & \gamma=2
  \end{array} \right.
\end{equation}
Fig.~\ref{mass-model}$b$ shows isocontours of the potential due to disk and
bulge.

In the future we plan to numerically evaluate the gaseous response to the mass
model and compare the resulting gas distribution and kinematics with those
observed (cf.\ the previous sections). Whether or not these data allow us to
pin down all the unknown quantities (e.g., the central mass concentration)
remains to be seen. However, the simulations published in the literature
suggest that gas observation do -- and thus {\em must\/} be used in order to --
constrain these quantities.

\section{Conclusions}
In the context of standard spiral density waves, the expected large-scale 
shocks lead to a phase shift between the gas response and driving stellar bar.
The resulting mass inflow may dominate standard viscous accretion. On much 
smaller scales, it is unlikely that such shocks can be sustained. Therefore, 
the shock mechanism is likely to be important at intermediate scales (0.1 to 
a few kpc). The mass inflow rate can then be computed in terms of the gas
properties, the underlying mass distribution, and the pattern speed.
In future we shall attempt a self-consistent fit to these parameters, 
particularly the shock strength, which will require gas dynamic modeling. 
Ultimately, we would like to address the implications of these observations 
for fueling nuclear activity. A mass model is a precursor to running a full 
hydro-dynamical simulation.

\acknowledgements
We thank H.~Liszt for allowing us to present unpublished VLA observations of
the \hi\ distribution, and B.~Garc\'{\i}a-Lorenzo who supplied the stellar
velocities for Fig.~\ref{gasrot}. We would also like to thank H.~Thronson for
the large-scale $K$ band image.

\end{document}